\documentclass[aps,prl,preprint,superscriptaddress]{revtex4-1}

\usepackage{graphicx}
\usepackage{hyperref}

\begin{document}

\title{Solid-state ensemble of highly entangled photon sources at rubidium atomic transitions}

\author{Robert Keil}
\email{R. K. and M. Z. contributed equally to this work}
\affiliation{Institute for Integrative Nanosciences, IFW Dresden, Helmholtzstra{\ss}e 20, 01069 Dresden, Germany}

\author{Michael Zopf}
\email{R. K. and M. Z. contributed equally to this work}
\affiliation{Institute for Integrative Nanosciences, IFW Dresden, Helmholtzstra{\ss}e 20, 01069 Dresden, Germany}

\author{Yan Chen}
\affiliation{Institute for Integrative Nanosciences, IFW Dresden, Helmholtzstra{\ss}e 20, 01069 Dresden, Germany}

\author{Bianca H{\"o}fer}
\affiliation{Institute for Integrative Nanosciences, IFW Dresden, Helmholtzstra{\ss}e 20, 01069 Dresden, Germany}

\author{Jiaxiang Zhang}
\affiliation{Institute for Integrative Nanosciences, IFW Dresden, Helmholtzstra{\ss}e 20, 01069 Dresden, Germany}

\author{Fei Ding}
\email[]{Correspondence should be addressed to F. D. : f.ding@ifw-dresden.de}
\affiliation{Institute for Integrative Nanosciences, IFW Dresden, Helmholtzstra{\ss}e 20, 01069 Dresden, Germany}
\affiliation{Institut f{\"u}r Festk{\"o}rperphysik, Leibniz Universit{\"a}t Hannover, Appelstra{\ss}e 2, 30167 Hannover, Germany}

\author{Oliver G. Schmidt}
\affiliation{Institute for Integrative Nanosciences, IFW Dresden, Helmholtzstra{\ss}e 20, 01069 Dresden, Germany}
\affiliation{Merge Technologies for Multifunctional Lightweight Structures, Technische Universit{\"a}t Chemnitz, Germany}

\date{\today}

\begin{abstract}
\textbf{Semiconductor InAs/GaAs quantum dots grown by the Stranski-Krastanov method are among the leading candidates for the deterministic generation of polarization entangled photon pairs~\cite{Benson2000,Stevenson2006,Akopian2006,Hafenbrak2007,Muller2009,Dousse2010}. Despite remarkable progress in the last twenty years, many challenges still remain for this material, such as the extremely low yield ($< 1\%$ quantum dots can emit entangled photons), the low degree of entanglement, and the large wavelength distribution. Here we show that, with an emerging family of GaAs/AlGaAs quantum dots~\cite{Rastelli2004} grown by droplet etching and nanohole infilling~\cite{Atkinson2012,Huo2013}, it is possible to obtain a large ensemble (close to $100\%$) of polarization-entangled photon emitters on a wafer without any post-growth tuning. Under pulsed resonant two-photon excitation, \emph{all measured} quantum dots emit single pairs of entangled photons with ultra-high purity, high degree of entanglement (fidelity up to $F=0.91$, with a record high concurrence $C=0.90$), and ultra-narrow wavelength distribution at rubidium transitions. Therefore, a solid-state quantum repeater - among many other key enabling quantum photonic elements - can be practically implemented with this new material.}
\end{abstract}

\maketitle

\section{Introduction} 
Solid-state sources that emit single pairs of entangled photons are a key element in quantum information technology. Polarization entangled photons from atomic cascades were firstly used to test Bell's inequality~\cite{Freedman1972,Aspect1981}, but demonstrating scalable applications with single atoms is clearly a technological challenge. In 1988 Shih and Alley reported that the photon pairs generated from spontaneous parametric down conversion (SPDC) are polarization entangled and can violate Bell's inequality~\cite{Shih1988,Burnham1970}, which opened the door for various polarization-entanglement based experiments. However, SPDC sources are characterized by Poissonian statistics, \emph{i.e.} one usually does not know when an entangled photon pair is emitted. This fundamentally limits their applications in complex quantum protocols.

Semiconductor InAs/GaAs quantum dots (QDs) grown by the Stranski-Krastanov method are among the leading candidates for the deterministic generation of polarization-entangled photons. As proposed by Benson \emph{et al.}, the cascaded emission in single QDs from the biexciton (XX, $|\uparrow\downarrow\Uparrow\Downarrow\rangle$) to the ground state via the intermediate exciton states (X, $|\uparrow\Downarrow\rangle$ or $|\downarrow\Uparrow\rangle$) produces polarization entangled photon pairs $|\psi^+\rangle=\frac{1}{\sqrt{2}}(|L_{XX}R_{X}\rangle+|R_{XX}L_{X}\rangle)$~\cite{Benson2000}, where $R$ and $L$ denote right- and left-handed circular polarization, respectively. In real InAs/GaAs semiconductor QDs, however, the anisotropy in strain, composition and shape reduces the QD symmetry and mixes the two bright exciton states, resulting in two non-degenerate bright exciton states $\frac{1}{\sqrt{2}}(|\uparrow\Downarrow\rangle\pm|\downarrow\Uparrow\rangle)$ split by the fine structure splitting (FSS)~\cite{Bayer2002}. The final two-photon state has a time-varying form $|\psi\rangle=\frac{1}{\sqrt{2}}(|HH\rangle+e^{iT_1{S}/\hbar}|VV\rangle)$, where $T_1$ is the radiative lifetime of the exciton and $S$ is the FSS~\cite{Stevenson2008}. In order to reduce the phase shift between the $|HH\rangle$ and $|VV\rangle$ two-photon components and to obtain a high degree of entanglement, the experimental strategies are to reduce the FSS $S$ and/or the exciton lifetime $T_1$.

This is unfortunately no easy task. In the last decade there have been extensive efforts to generate entangled photons with InAs/GaAs QDs. The probability of finding suitable QDs in an as-grown sample is $< 10^{-2}$~\cite{Juska2013,Gong2014}, thus necessitating the use of post-growth tuning techniques (such as thermal annealing, optical Stark effect, magnetic, electric and strain fields) to eliminate the FSS~\cite{Plumhof2012}. On one hand, the fact that every single QD needs to be independently engineered imposes a great challenge for the practical application of QD-based devices. On the other hand, due to the electron-nuclear spin hyperfine interactions~\cite{Welander2014,Hudson2007}, the degree of entanglement of InAs/GaAs QD-based sources is generally low even at zero FSS. The best result so far yields an entanglement fidelity $F=0.82$ and concurrence $C=0.75$~\cite{Trotta2014}. Alternatively one can reduce the exciton lifetime $T_1$ by using the Purcell enhancement in a cavity, or perform a time gating before a significant phase shift $T_1{S}/\hbar$ between $|HH\rangle$ and $|VV\rangle$ takes place. The former requires a simultaneous Purcell enhancement of both X and XX emissions~\cite{Dousse2010}, which is a non-trivial task. The latter discards a large portion of photons and reduces the source brightness significantly.

Is it possible to obtain a large ensemble of QDs emitting entangled photons, without using any post-growth tuning? The answer is yes, if one can grow QDs with highly symmetric confinement, short lifetime and ideally, weak electron-nuclear spin hyperfine interactions. The first attempt was reported by Juska \emph{et al.}~\cite{Juska2013}, where arrays of symmetric In$_{0.25}$Ga$_{0.75}$As$_{1-\delta}$N$_{\delta}$ were grown on the GaAs (111)B surface. They were able to obtain areas with an impressive 15\% of entangled photon emitters with a fidelity $F>0.5$. Although the FSS is consistently below 4 $\mu eV$ for these novel QDs, the exciton lifetime is quite long ($1.8\pm0.6~ns$). Kuroda \emph{et al.}~\cite{Kuroda2013} demonstrated the generation of entangled photons (with fidelity up to $F=0.86$) using highly symmetric GaAs QDs grown on the GaAs (111)A surface by droplet epitaxy. Although the exciton lifetime is short ($560~ps$), the FSS are relatively large (with a mean value of $10\pm5~{\mu}eV$) and the hyperfine interaction of the exciton with nuclear spins is significant in this system~\cite{Kuroda2013}.

Here we show, for the very first time, that a large ensemble of \emph{as-grown} polarization-entangled photon emitters can be obtained, by using an emerging family of GaAs/AlGaAs QDs grown by droplet etching and nanohole infilling. These QDs exhibit very small FSS (with a mean value of $4.8\pm2.4~{\mu}eV$), short lifetime ($T_1 < 220 ps$) and ultra-narrow wavelength distribution at rubidium transitions. Under pulsed resonant two-photon excitation, pronounced Rabi oscillations can be observed up to 7$\pi$ and \emph{all measured} QDs emit single pairs of entangled photons with ultra-high purity and high degree of entanglement (fidelity $F$ up to 0.91, with a record high concurrence $C$ = 0.90). We envision that a number of key enabling quantum photonic elements can be \emph{practically} implemented by using this novel material system. A particularly important example is a hybrid quantum repeater, where the QD-generated entangled photon qubits can be mapped reversibly in and out of a rubidium vapor based quantum memory.

\section{Results and discussion}
\subsection{Sample growth}

\begin{figure}[!h]
\centering
\includegraphics[width=0.95\textwidth]{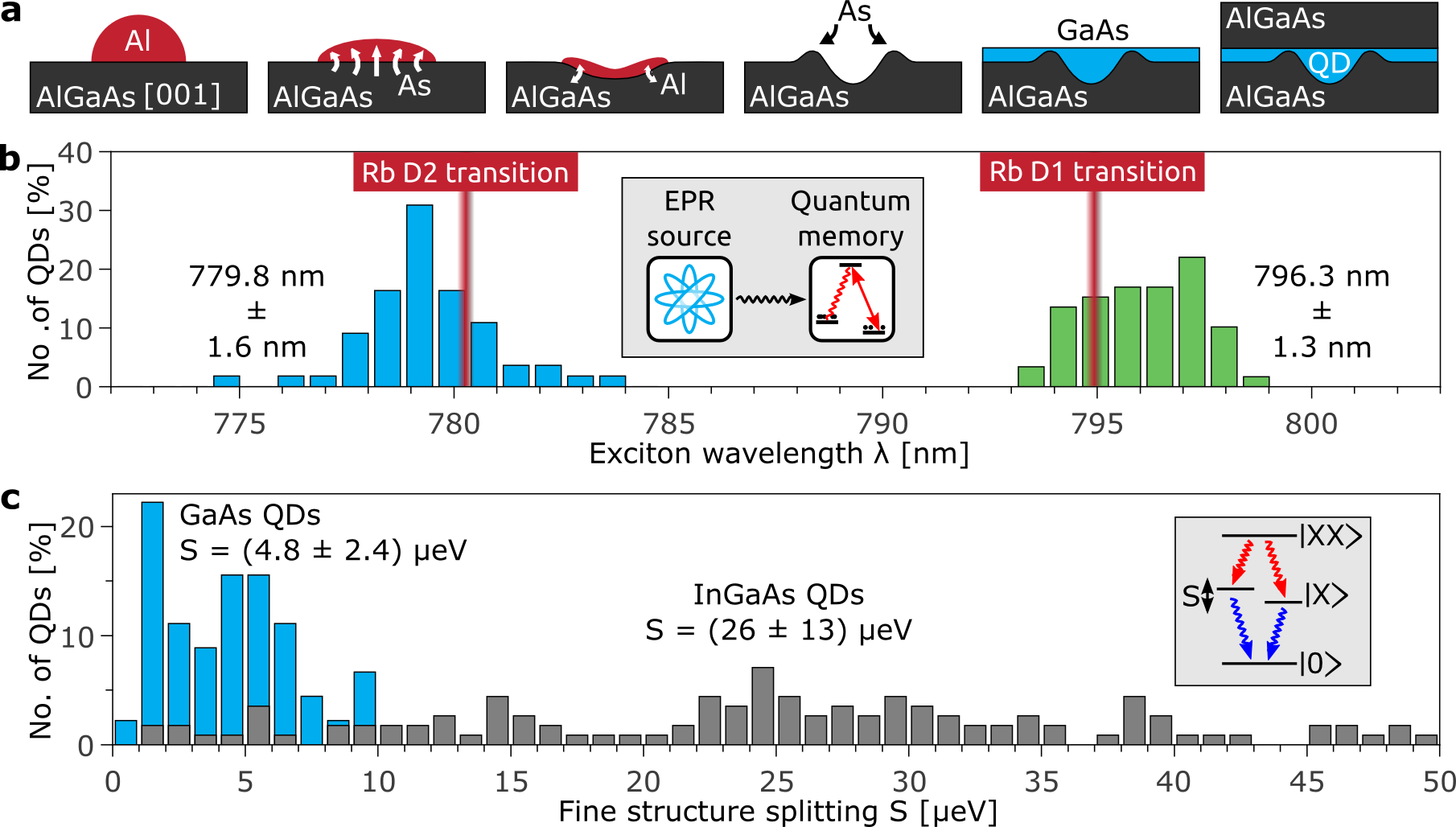}
\caption{\textbf{Growth and properties of highly homogeneous GaAs/AlGaAs quantum dots.} \textbf{a}, Processing steps during the growth of GaAs/AlGaAs quantum dots. \textbf{b}, Exciton emission wavelength distribution for two different samples with GaAs infilling amounts of 2 nm (blue) and 2.75 nm (green) for more than 50 dots measured on each sample. Red lines indicate the rubidium D1 and D2 transition at 794.9 and 780.2 nm, respectively. Inset: sketch of envisioned interface between entangled photons from a QD and an atomic quantum memory based on the Raman scheme\cite{Ding2015}. \textbf{c}, Occurrence of the exciton fine structure splitting, comparing the GaAs/AlGaAs QDs (blue) with InAs/GaAs QDs (grey). Inset: scheme of the biexciton (XX) decay indicating the spin related fine structure splitting $S$ between the intermediate exciton states (X). }
\label{fig1}
\end{figure}

The QDs presented in this work are fabricated by solid-source molecular beam epitaxy. In-situ droplet etching is utilized to create self-assembled nano-holes with ultra-high in-plane symmetry~\cite{Atkinson2012,Huo2013}, which are subsequently filled and capped to obtain embedded solid-state quantum emitters~\cite{Rastelli2004}. \autoref{fig1}a shows a sketch of the processes involved in the QD formation. Initial point is a GaAs (001) substrate which has been deoxidized and overgrown with a GaAs buffer layer followed by 200 nm of Al${_x}$Ga$_{1-x}$As. First, Al is deposited under low arsenic pressure ($< 10^{-8}$ mbar), forming droplets on the surface at $630$ degree. Driven by concentration gradients, the concurring dissolution of As through the droplets and diffusion of Al towards the substrate induces the formation of nano-holes with high symmetry. In a following annealing step the structures crystallize under a re-established As atmosphere. Then the nano-holes are filled with GaAs and subsequently overgrown by Al${_x}$Ga$_{1-x}$As to obtain the isolated QDs with three-dimensional carrier confinement.

Envisioning a hybrid QD-atomic interface as a promising solid-state quantum memory~\cite{Akopian2011,Pan2012}, it is desirable to match the QD emission with atomic transitions, illustrated by the inset in \autoref{fig1}b. For this purpose several samples with varying GaAs infilling amounts have been grown, targeting the Rb D1 and D2 transition line at a wavelength of 794.9 nm and 780.24 nm, respectively. \autoref{fig1}b shows the exciton wavelength distribution for two different samples with 2 nm (blue) and 2.75 nm (green) GaAs deposited at a growth rate of 0.47 and 0.5 ${\mu}m/h$, accordingly. The statistics on more than 50 QDs across a large area on each sample show an unprecedented control on the central emission wavelengths, with mean values of ($779.8\pm1.6$) nm and ($796.3\pm1.3$) nm. The wavelength distributions, or the so-called inhomogeneous broadenings, are by far the narrowest for semiconductor QDs and are about 5 times narrower than that of a typical self-assembled InAs/GaAs QD sample.

Together with the high homogeneity, the QDs also exhibit high symmetry due to the negligible intermixing and a virtually strain free interface between GaAs and AlGaAs. Previous work suggests that a reduction of the amount of deposited Al and an increase of the deposition rate can enhance the nano-hole symmetry~\cite{Huo2013}. Following this trend, a single pulse of 0.09 nm excess Al at a growth rate of 0.8 ${\mu}m/h$ (corresponding to AlAs growth) was used for our samples. The optimized growth protocols lead to a very satisfying result. \autoref{fig1}c shows the statistical distribution of the FSS for the GaAs/AlGaAs QD sample studied in this work (blue) and for a typical InAs/GaAs QD sample grown by partial capping and annealing (grey). The total number of measured dots is 45 and 114, respectively. The GaAs QDs feature an average FSS of only ($4.8\pm2.4$) $\mu$eV, which is among the best reported so far\cite{Huo2013,Kuroda2013,Juska2013}. With these superior spectral properties, the investigated samples are promising candidates for the generation of polarization entangled photons.

\subsection{Resonant excitation of the biexciton}
\begin{figure}[!h]
\centering
\includegraphics[width=0.85\textwidth]{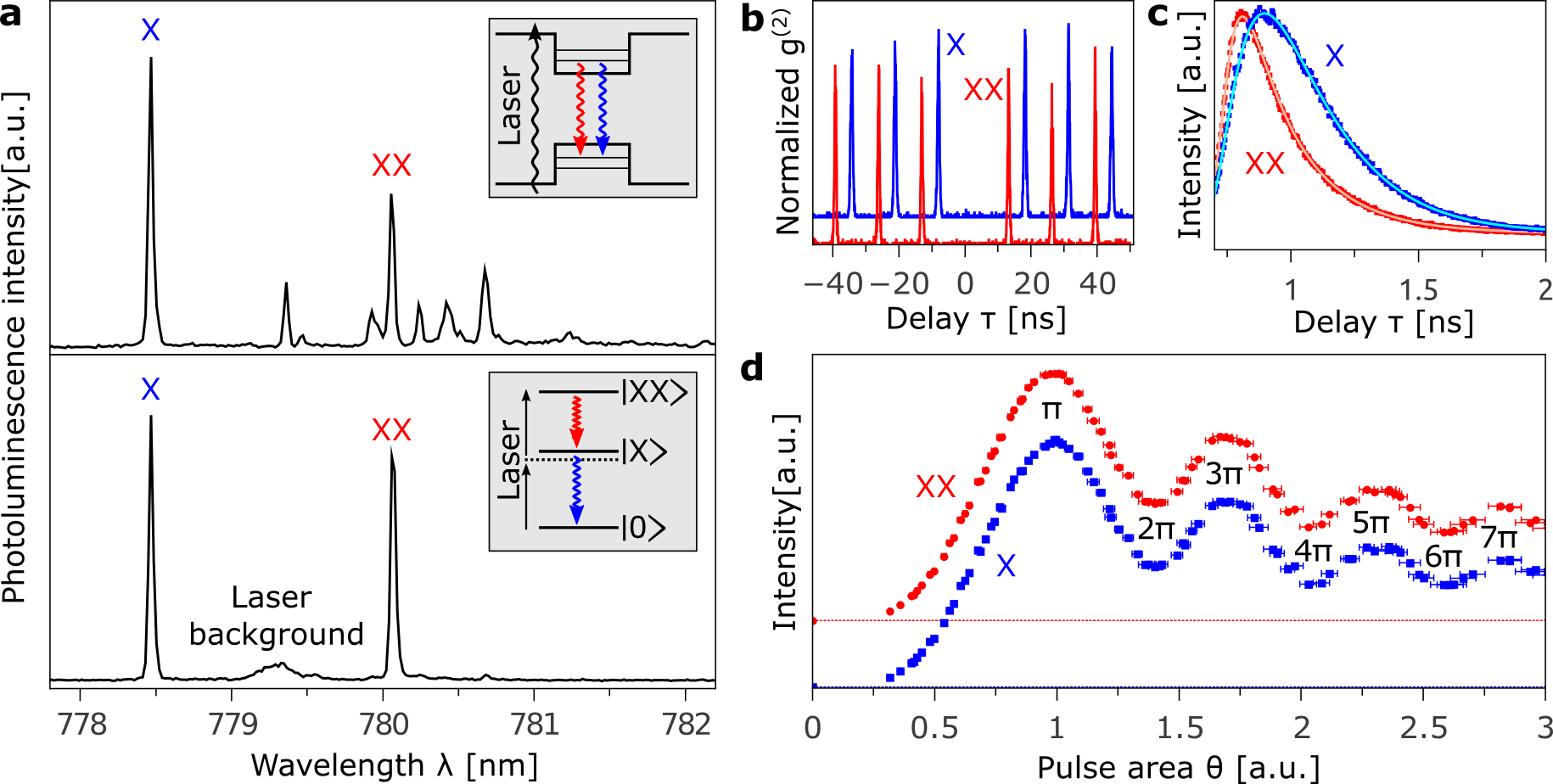}
\caption{\textbf{Coherent excitation of the XX transition in GaAs/AlGaAs QDs using a two-photon excitation scheme.} \textbf{a}, QD emission spectrum for pulsed above-band excitation (top) and resonant XX excitation (bottom). Non-resonant excitation reveals the dominant X transition and the XX transition which are spectrally close to other excitonic species. In order to efficiently and coherently pump the XX state, a two-photon excitation scheme is employed which results in a pure spectrum showing the XX-X cascade. \textbf{b}, Intensity-Autocorrelation measurement of the XX and X transition showing the normalized coincidences plotted over the delay time $\tau$. Very pure single photon emission is confirmed by: $g^{(2)}_{XX} = 0.02 \pm 0.02$ and $g^{(2)}_{X} = 0.04 \pm 0.02$. \textbf{c}, Measurements of the fluorescence lifetime $T_1$. Short radiative lifetimes of $T_{1,XX}$ = 112 ps and $T_{1,X}$ = 134 ps are determined. \textbf{d}, Rabi oscillations of the XX and X emissions as a function of the pulse area $\Theta$ }
\label{fig2}
\end{figure}

The major challenge to realize entangled photon emissions from GaAs/AlGaAs QDs is the effective excitation of the biexciton (XX) state. So far there are only very few reports about the observation of a biexciton \cite{Ghali2012,Kuroda2013,Graf2014} in GaAs/AlGaAs QDs (although, grown with different methods) due to the low internal population probability under non-resonant excitation. Due to the optimized growth process we are able to observe strong XX emissions even with above-band excitations. We select a QD from the sample emitting close to the Rb D2 transition ($\sim$780.2 nm) and excite it by pumping the surrounding higher-bandgap AlGaAs with a pulsed laser. The resulting spectrum (\autoref{fig2}a: top), which is relatively clean in a broad range, shows several different excitonic transitions: The transition with the highest intensity is the exciton emission (X) at $\lambda$ = 778.5 nm. Among several red-shifted transitions the XX emission is the strongest ($\lambda$ = 780.1 nm).

In order to efficiently excite and to coherently drive the XX transition, we pump the two-photon resonance of the XX state by using a strong laser which lies spectrally in between the X and XX transitions. This excitation scheme has already been proven very effective in case of InAs/GaAs QDs \cite{MullerM2014} and has recently been reported by Huber et al~\cite{Huber2016} in case of selected GaAs/AlGaAs QDs also grown by local droplet etching. Making use of tunable notch filters we can effectively suppress the laser background. Hence a very pure spectrum showing mostly the XX and X emissions can be observed (\autoref{fig2}a: bottom). The integrated intensities are the same for both emissions, strongly indicating a close to unity efficiency for the cascaded emission process~\cite{MullerM2014}.

To obtain the evidence of pure single photon emissions from both XX and X, we perform an autocorrelation measurement using a standard Hanbury Brown and Twiss setup and the results are shown in \autoref{fig2}b. The autocorrelation function $g^{(2)}(\tau)$ plotted over the photon arrival delay $\tau$ shows a clear absence of counts at zero delay and proves the ultra-high purity single photon emission. The background-corrected correlation function is measured to be $g^{(2)}_{XX} = 0.02 \pm 0.02$ for XX and $g^{(2)}_{X} = 0.04 \pm 0.02$ for X.

Next, we measured the luminescence lifetime $T_1$ by recording an intensity correlation between the excitation laser pulse and the arrival time of the photons, see \autoref{fig2}c. The XX shows a simple exponential decay which is fitted taking into account the convolution with the detector response function. The X decay shows a longer rise time since the state has to be populated first by the decay of the XX state. The extracted lifetimes are $T_{1,XX}$ = 112 ps and $T_{1,X}$ = 134 ps, which are among the lowest values recorded for as-grown semiconductor QDs. The ideal lifetime-limited linewidth of the exciton emission is therefore $\Delta E = 4.9 \mu eV$, which is close to the mean value of the FSS in our sample.

In order to further evaluate the resonant two-photon excitation scheme we recorded the intensity of XX and X photons while changing the excitation power. The result is summarized in \autoref{fig2}d by plotting the intensity over the pulse area $\Theta$ which is proportional to the square root of the excitation power. Clear Rabi oscillations are observed, which are oscillations of the intensity due to a coherent rotation on the Bloch sphere between the ground state $|0>$ and the excited state $|XX>$. The abscissa is normalized in units of $\pi$ to the first maximum of the XX intensity, where the pulse area is equal to $\pi$. Intensity oscillations up to $7\pi$ are observed. The mean intensity is decreasing for higher excitation powers, which may be caused by several different factors, like chirp in the excitation pulse or scattering processes in the QD~\cite{MullerM2014}. Increasing the power also leads to an increase in the oscillation frequency, which is a fingerprint of the two-photon excitation process in clear contrast to one-photon resonant excitation, where the frequency remains constant.

\subsection{Evaluating the degree of entanglement}
After realizing an efficient coherent control over the XX decay in GaAs QDs we now evaluate the degree of entanglement in the polarization of the emitted photons. A QD with a FSS of $S=2.3\mu eV$ is chosen in the experiment, since it presents a large portion ($\sim 22\%$) of QDs in the sample (see \autoref{fig1}c). The QD is excited with $\pi$-pulses for an efficient preparation of biexciton states. To measure the degree of polarization correlation we send the stream of XX and X photons on a $50:50$ beam splitter. Each subsequent signal arm contains a quarter-wave plate, a half-wave plate and a polarizer in order to select the polarization in an arbitrary basis. After spectral selection of XX and X photons in the first and second signal arm, respectively, they are sent to single photon detectors. Coincidence counting hardware is used to obtain the second-order correlation function $g^{(2)}_{XX,X}$ between XX and X photons for the selected polarization direction.

\begin{figure}[!h]
\centering
\includegraphics[width=0.95\textwidth]{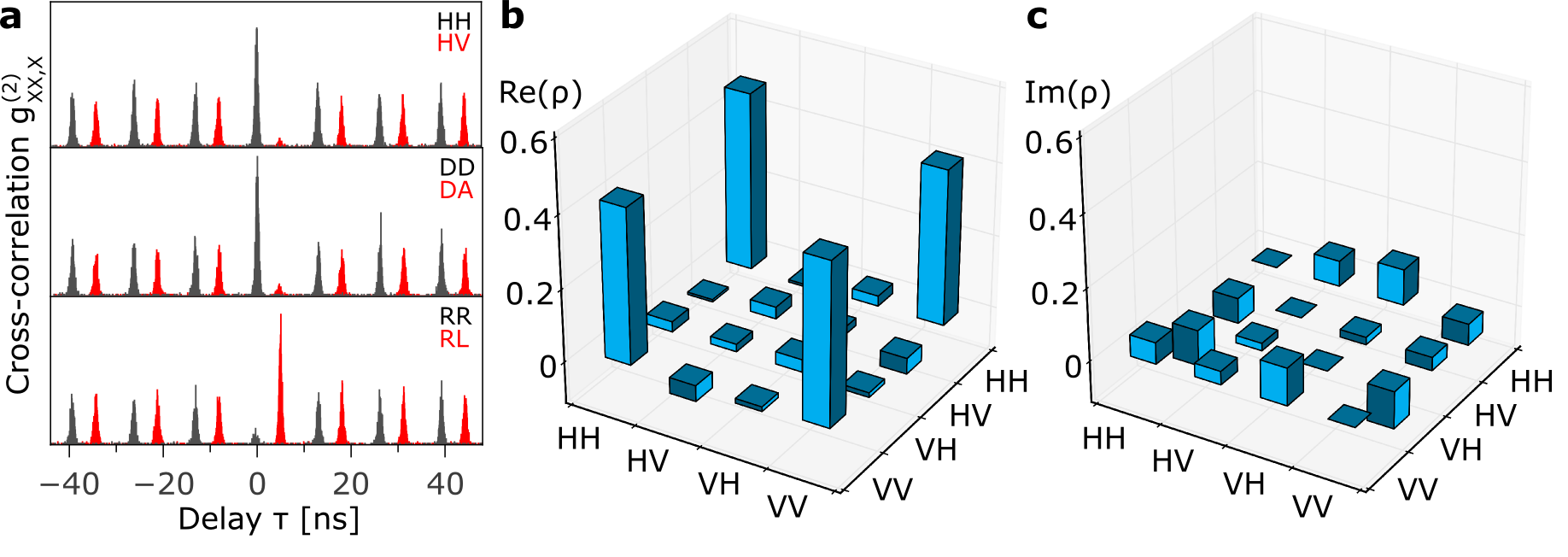}
\caption{\textbf{Degree of Entanglement from a QD with finite FSS.} \textbf{a} Cross-correlation measurements between XX and X photons for co- and cross-polarized photons in the rectilinear (H,V), diagonal (D,A) and circular (R,L) bases (top to bottom). From these measurements a fidelity $F = 0.88 \pm 0.03$ to the state $|\psi^+\rangle$ is deduced. \textbf{b-c} Two-photon density matrix reconstructed from 16 correlation measurements on the same QD, employing the maximum likelihood technique, \textbf{b} showing the real and \textbf{c} the imaginary part. The fidelity and concurrence extracted from this matrix are $F = 0.91$ and $C = 0.90$, respectively.}
\label{fig3}
\end{figure}

\autoref{fig3}a shows 6 cross-correlation measurements obtained for three bases of co-polarized and cross-polarized photons: the recilinear (HV), diagonal (DA) and circular (RL) basis (from top to bottom). As expected for an ideal entangled two-photon state $|\psi^+\rangle=\frac{1}{\sqrt{2}}(|L_{XX}R_{X}\rangle+|R_{XX}L_{X}\rangle)$, a strong bunching (antibunching) at $\tau=0$ is observed for co-polarization (cross-polarization) in the rectilinear and orthogonal bases, whereas this behavior is reversed for the circular basis set. The correlation contrast for a chosen basis set is given by\cite{Hudson2007}
\begin{equation}
C_{basis} = \frac{g^{(2)}_{XX,X}(0) - g^{(2)}_{XX,\bar{X}}(0)}{g^{(2)}_{XX,X}(0) + g^{(2)}_{XX,\bar{X}}(0)}
\end{equation}
with $g^{(2)}_{XX,X}(0)$ denoting the second-order correlation function at zero delay in collinear, and $g^{(2)}_{XX,\bar{X}}(0)$ in orthogonal bases. For the three illustrated basis sets the following contrasts are obtained:
\begin{eqnarray}
C_{linear} = 0.89 \pm 0.03 \\
C_{diagonal} = 0.83 \pm 0.04 \\
C_{circular} = -0.78 \pm 0.04
\end{eqnarray}
The fidelity F of the measured quantum state to the ideal state $|\psi^+\rangle$ can then be obtained by\cite{Hudson2007}
\begin{equation}
F = \frac{1 + C_{linear} + C_{diagonal} - C_{circular}}{4} = 0.88 \pm 0.03
\end{equation}
which exceeds the classical limit $F=0.5$ by more than 12 standard deviations.

A more comprehensive picture of the measured entangled two-photon state is given by the density matrix representation. We performed cross-correlation measurements for 16 different base combinations to account for the 16 unknown variables in the density matrix $\rho$. The measured values for $g^{(2)}(0)$ are then used to construct a density matrix following the procedure presented in Ref.~\onlinecite{James2001}. Since the thereby obtained density matrix violates important basic properties like positive semidefiniteness, the maximum likelihood estimation is employed to find the appropriate density matrix which is the closest to the measured results. The resulting matrix is shown in \autoref{fig3}, split into the real part (\autoref{fig3}b) and imaginary part (\autoref{fig3}c). The strongest features are observed in the outer-diagonal real-part matrix elements, which are close to $0.5$, while all other elements are close to zero. This is in agreement with the expected entangled state $|\psi^+\rangle$ whose density matrix should have only non-zero values of $0.5$ in the outer-diagonal elements. The small (but non-zero) real values in the off-diagonal elements indicate a weak spin scattering process in the QD. The finite imaginary off-diagonal values represent a small phase difference between $|HH\rangle$ and $|VV\rangle$, which may be caused by the joint effect of a finite FSS and an accumulated phase due to the optical setup. From this density matrix, we obtain a fidelity F (after background corrections~\cite{MullerM2014}) to the state $|\psi^+\rangle$ of
\begin{equation}
F = \langle\psi^+| \rho |\psi^+\rangle = 0.91
\end{equation}
which is very close to the value of $0.88\pm0.03$ obtained from the 6 cross-correlation measurements in \autoref{fig3}a.

Another measure for non-classical properties of a quantum state is the Concurrence $C$ \cite{James2001}. Using the acquired density matrix, a value of $C=0.90$ (raw data without correction: $C=0.81$) is obtained, which is not only the best value measured for InAs/GaAs QDs with zero FSS~\cite{Trotta2014}, but also the highest value ever obtained for any QD entangled photon source. The high values for fidelity and concurrence are especially remarkable considering the finite fine structure splitting of $S=2.3\mu eV$, which already significantly degrades the entanglement in case of InAs/GaAs QDs \cite{Zhang2015b,Hudson2007}.

\begin{figure}[!h]
\centering
\includegraphics[width=0.5\textwidth]{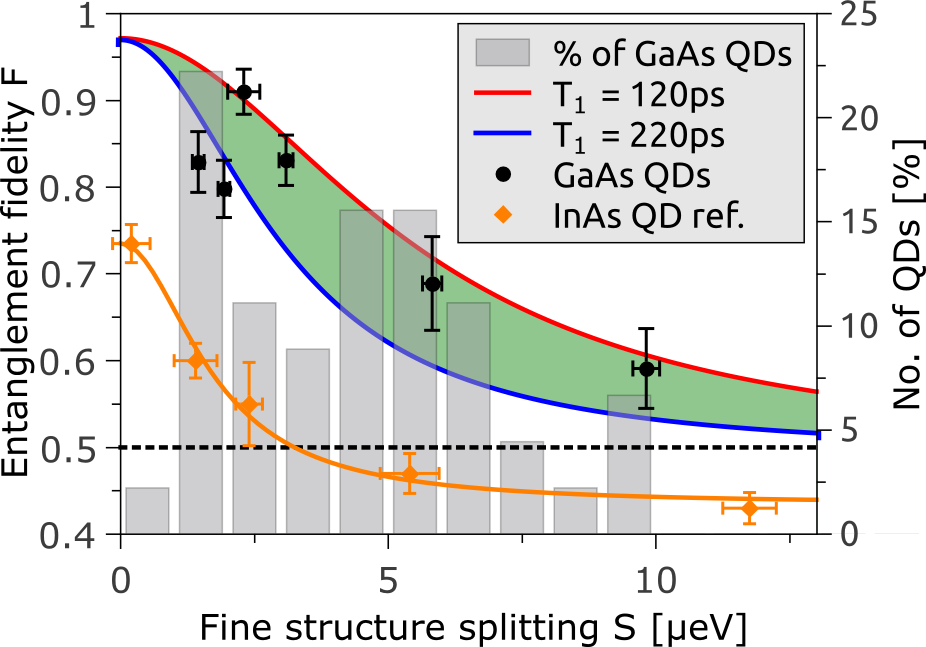}
\caption{\textbf{Entanglement fidelity of GaAs/AlGaAs and InAs/GaAs QDs as a function of FSS.} All measured QDs (black circles) from the sample emit entangled photons with a fidelity above the classical limit of $F=0.5$ (dashed line). The highest fidelity is measured to be $F = 0.91$, and even the QD with $S=9.8\mu eV$ shows $F=0.59$. Fidelity values of InAs/GaAs QDs taken from Ref.~\onlinecite{Zhang2015b} are plotted in orange, together with a Lorentzian fit. A theoretical model of the fidelity is plotted for radiative lifetimes $T_1$ = 120 ps (red) and $T_1$ = 220 ps (blue), which represents the range of all measured values for $T_1$ in our sample. Together with the FSS distribution (grey histogram, as also in \autoref{fig1}c), the fidelity measurements strongly indicate that close to 100\% of the QDs in the sample are entangled-photon emitters. }
\label{fig4}
\end{figure}

Since the phase shift $T_{1}{S}/\hbar$ between $|HH\rangle$ and $|VV\rangle$ states is significantly reduced due to the very short lifetime $T_1$ in this system, we expect that the generation of entangled photons should be also possible for QDs with even higher values of $S$. Therefore we select six dots representing the whole range of FSS measured in the sample. By measuring six cross-correlations in three basis sets for each dot, their entanglement fidelity $F$ are obtained. \autoref{fig4} shows the values of $F$ plotted as a function of FSS (black circles), overlaid on the FSS distribution in the sample (gray histogram).
The data from Zhang \emph{et al}. \cite{Zhang2015b} including a Lorentzian fit are shown as a reference for typical InAs/GaAs QDs (orange line). Remarkably, all of the measured dots show clear signature of entangled photon emission with $F>0.5$. Even the QD with $S=9.8\mu eV$, which represents the QDs with the largest FSS in our sample, shows a fidelity $F=0.59 \pm 0.05$. We want to highlight that the measured dots were not preselected according to certain conditions apart from their FSS. All measurements lead to the conclusion that nearly 100\% of the QDs in this sample show entangled photon emission.

Another outstanding feature of the new system are the significantly higher fidelities compared to that of the typical InAs/GaAs QDs, which is mostly due to the weak electron-nuclear spin hyperfine interactions in this type of QDs~\cite{Welander2014,Hudson2007,Ulhaq2016}. In order to better understand the obtained values, we plot two theoretical curves showing the fidelity over the FSS for radiative lifetimes of $T_1$ = 120 ps (red curve) and $T_1$ = 220 ps (blue curve), which is the typical range for the measured lifetimes. We modelled the fidelity following the work by Hudson \emph{et al}. \cite{Hudson2007}, which includes both the influence of the FSS and lifetime $\tau$ as well as cross-dephasing and spin scattering:
\begin{equation}
F = \frac{1}{4} \left( 1 + k g^{'(1)}_{H,V} + \frac{2kg^{(1)}_{H,V}}{1+x^2} \right)
\end{equation}
with
\begin{equation}
x = \frac{g^{(1)}_{H,V} S T_{1,X}}{\hbar}
\end{equation}
Here, $k$ denotes the probability that the measured photon pairs originate from the dot, which we estimate to be $k=0.97$ due to the measured autocorrelation measurements presented in \autoref{fig2}b. The factor $g^{(1)}_{H,V}$ denotes the fraction of the QD emission which is unaffected by both cross-dephasing and spin-scattering processes, while $g^{'(1)}_{H,V}$ only considers spin-scattering processes. Since the presented data show no trend that would lead to $F<0.5$ for large FSS, we expect spin-scattering processes to have a negligible influence on the entanglement. Considering spin scattering due to the Overhauser field of the nuclear spins present in the dot, a spin-scattering time of $T_{SS}=15ns$ can be assumed \cite{Chekhovich2013}. This is, however, two orders of magnitudes longer than our measured radiative lifetimes and therefore barely contributes in the degradation of the fidelity. On the other hand, in typical InAs/GaAs QDs this effect can be significantly stronger in case of high concentrations of spin-9/2 Indium \cite{Chekhovich2013}, leading to much lower fidelity values for InAs/GaAs QDs in \autoref{fig4}~\cite{Zhang2015b}. Since the fidelity at small FSS are very high for the GaAs/AlGaAs QDs, we neglected cross-dephasing processes in the model. It is clear that the trend in all our fidelity measurements can be well-represented by the employed model.

\section{Conclusion}
In this work we propose a new type of solid-state polarization entangled photon source based on an emerging family of GaAs/AlGaAs QDs. These QDs can be grown with unprecedented wavelength control, ultra-small FSS and short lifetime. The efficient and coherent excitation of the biexciton state in the GaAs/AlGaAs QDs is achieved by employing a resonant two-photon excitation scheme. A fidelity up to $F=0.91$ and a concurrence of $C=0.90$ have been achieved, which are among the highest values ever reported for QD-based entangled photon sources. Most remarkably, the whole set of measurements draws an unambiguous conclusion that we have obtained a large ensemble of entangled photon emitters on a single semiconductor wafer. With almost 100\% of QDs in the sample having fidelities $F>0.5$, a great fraction of QDs are expected to exhibit high fidelities $F>0.8$ without any post-growth tuning. Therefore, a solid-state quantum repeater - among many other key enabling quantum photonic elements based on polarization entangled photons - can be \emph{practically} implemented with this new material system.

\section{Methods}
The photoluminescence experiments were conducted at $T=4K$ by placing the sample in either a helium-bath or a helium-flow cryostat. As excitation laser for the above-band and two-photon excitations a pulsed Ti:Sa laser with $76MHz$ repetition rate was used, which generated pulses with a duration of $3ps$. In order to spectrally narrow the laser pulse it was sent to a home-built pulse-shaping setup before it was coupled into a single-mode fiber. The excitation laser was then sent to the sample in the cryostat using a beam sampler and focussed by a lens or an objective, which was also used for the collection of the QD emission. We used half-ball solid immersion lenses to increase the photon collection from the sample. For measuring entanglement the collected light from the QD was split by a $50:50$ beam splitter into two arms, each consisting of a quarter-wave plate, a half-wave plate and a polarizer. The two beams were then coupled into polarization-maintaining single-mode fibers. In order to eliminate the strong laser background we employed two consecutive tunable notch filters. Each light path was fed into a monochromator in order to select the XX or X transition, respectively. The streams of photons were then detected by avalanche photodiodes, whose signals were processed by a time-correlated single photon counter.
We measured the FSS of the sample by rotating the half-wave plate in the entanglement setup by $\alpha$ while rotating the quarter-wave plate by $2\alpha$. By obtaining high-resolution spectra for multiple values of $\alpha$ we were able to fit the emission lines and determine an oscillation amplitude of the peaks spectral center position which corresponds to the FSS. The FSS in the InAs/GaAs QD reference sample was determined using the same method, but simply by rotating a half-wave plate in front of a linear polarizer.
The lifetimes $T_1$ were obtained using an avalanche photodiode with a short response function, which was measured by using 3 ps laser pulses to be FWHM$\approx$ 100 ps. The measured response function was used to obtain the convoluted theoretical fits.

\section{Acknowledgments}
The work was financially supported by the BMBF Q.Com-H (16KIS0106) and the European Union Seventh Framework Programme 209 (FP7/2007-2013) under Grant Agreement No. 601126 210 (HANAS). The authors thank E. Zallo, N. Akopian, K. J{\"o}ns, A. Rastelli for helps and fruitful discussions, and Y. Li, X. Zhang, B. Eichler, R. Engelhard, S. Nestler, S. Baunack and S. Harazim for technical supports.

%

\end{document}